\documentclass[showpacs,twocolumn,prb,aps]{revtex4}
\usepackage{epsfig}
\usepackage{amsmath}
\usepackage{amssymb}
\begin{document}
\title{Quantum Monte Carlo simulation for the conductance of one-dimensional 
       quantum spin systems}

\author{Kim Louis and Claudius Gros} 

\affiliation{Fakult\"at 7, Theoretische Physik,
 University of the Saarland,
66041 Saarbr\"ucken, Germany.}

\date{\today}

\begin{abstract}
Recently, the stochastic series expansion (SSE) has been proposed
as a powerful MC-method, which allows simulations at low $T$
for quantum-spin systems. We show that the SSE allows to
compute the magnetic conductance for various one-dimensional
spin systems without further approximations.
We consider various modifications of the anisotropic Heisenberg
chain. We recover the Kane-Fisher scaling for one impurity
in a Luttinger-liquid and study the influence of non-interacting
leads for the conductance of an interacting system.
\end{abstract}
\maketitle

\section{Introduction}

In general, integrable one-dimensional models  show an ideally
conducting behavior in contrast to most real three-dimensional materials.
\cite{Cas95} There has  been intensive investigation of the
influence of the integrability on the conductivity for various
model systems.\cite{Cas95,Zot96,Zot97,Alv02} The conductance
of conducting, nearly one-dimensional devices is, on the
other hand, of substantial experimental interest.
Over the last years it has become possible
to fabricate mesoscopic devices,\cite{Imry} 
such as carbon nanotubes which can be viewed as a realization 
of systems with ballistic transport properties.
Therefore, the computation of dynamical transport quantities
has received considerable interest.

A basic approach for the study of the 
conductance has been in the past via the bosonization of
appropriate model systems,\cite{ApelRice,KaneFish,GogNerTsve}
valid in the low-temperature limit. Numerical studies
have  so far involved the density-matrix-renormalization-group
(DMRG) technique \cite{Marston,Molina}
and Monte Carlo (MC).\cite{Moon,Leung}
In the case of Ref.\ \onlinecite{Marston} a reduced set of states
was used in order to evaluate the dynamics, in
Ref.\ \onlinecite{Molina}, a phenomenological formula
by Sushkov\cite{Sushkov} was used to compute the conductance
(see also Ref.\ \onlinecite{Mescholl}). 
The simulations by Refs.\ \onlinecite{Moon,Leung} use an effective
bosonized Hamiltonian as a starting point.

Here we will discuss how to obtain the conductance with 
quantum-Monte-Carlo (QMC) on
the original lattice Hamiltonian. For this purpose the
conductance will be calculated on the imaginary frequency axis.
We will show that a reliable extrapolation to zero frequency
can be performed at finite but low temperatures.
We will thus obtain 
approximation-free results for the dynamics of inhomogeneous
quantum-spin systems at low but finite temperatures, within a well
defined numerical accuracy defined by the statistics of the
MC-sampling and the accuracy of the zero-frequency extrapolation.

By the Jordan-Wigner transform a one-dimensional 
spinless fermionic system can be mapped to a
hard-core boson model.
Hence, it is possible to calculate the conductance for a fermionic
system in a bosonic one. This is vitally important since boson
models can be easily analyzed by Monte Carlo simulations where the sign
problem is absent.
However, for an evaluation of the conductance one requires a highly
efficient simulation method
which performs well at low temperatures.
Recently,\cite{Sandvik} such a  powerful method has been proposed: the 
Stochastic Series Expansion (SSE).
In this paper we will compute the conductance in a hard-core boson
lattice model by the aid of this new method.

\section{explanation of the method }
\subsection{Definition of the Conductance}\label{subSec_def_g}

We consider the anisotropic $xxz$-Hamiltonian
$$H_{xxz}=\sum_{n=1}^{N-1}
J_x\left(S_n^+S_{n+1}^-+S_n^-S_{n+1}^+\right)/2+J_zS_n^zS_{n+1}^z,
$$
where the $S_{n}^\pm=S_n^x\pm i S_n^y$ are the raising/lowering
operators for spin-1/2 Heisenberg spins.
The spin-current operator $j_n$ at a given site $n$ follows from
the continuity equation and is given by (see e.g.\ Ref. \onlinecite{Alv02})
$$j_n=iJ_xe\left(S_n^+S_{n+1}^--S_n^-S_{n+1}^+\right)/(2\hbar).$$
As a perturbation we will use a local ``voltage drop''. In the hard-core
boson notation this corresponds to a step in chemical potential---at
site $m$--- which is
equivalent to
$$P_m=e\sum_{n>m}S^z_n,$$
in terms of the Heisenberg-spin operators.

The conductance is then defined as the dynamical response of
the current operator at site $x$ to the voltage drop at site $y$:
\begin{equation}\label{defg}
g:=\lim_{z\to 0}{\rm Re}\frac{i}{\hbar}\int_0^\infty e^{izt}\langle
[j_x(t),P_y]\rangle \,dt.
\end{equation}
For open boundary conditions (OBC)
the relation $i[H,P_x]=\hbar\, j_x$ holds and a 
partial integration of (\ref{defg}) yields:
\begin{eqnarray*}
g
&=&\!{\rm Re}\left[(-iz)^{-1}\frac{i}{\hbar}\Bigl\{\langle [j_x,P_y]\rangle -
\int_0^\infty\!\! e^{izt}\langle[j_x(t),j_y]\rangle dt\Bigr\}\right]. \\
\end{eqnarray*}
Using ${\rm Re}(ab)={\rm Re}a{\rm Re}b-{\rm Im}a{\rm Im}b$
in the  above
equation gives two contributions to the conductance.
With the definition of the generalized Drude weight
 for two operators $A$ and
$B$:
%
$$
\langle\langle AB\rangle\rangle \ \equiv\ 
\lim_{z\to 0}(-iz)\int_0^\infty
e^{itz}\langle \Delta A(t)\Delta B\rangle\, dt~.
$$
where $\Delta A=A-\langle A\rangle$ 
one can show using the Lehmann representation\cite{louis}
that
the first  contribution (${\rm Re}a{\rm Re}b$) reads
\begin{equation}
\langle\langle j_xj_y\rangle\rangle\,{\rm Re}(-iz)^{-1} =
\langle\langle j_xj_y\rangle\rangle\,\pi\,\delta({\rm Re}z).
\label{the_first_term}
\end{equation}

The second factor in the expression (\ref{the_first_term})
[namely, ${\rm Re}(-iz)^{-1}$] gives rise to a delta-function,
 such that (\ref{the_first_term}) should
be the dominating contribution to $g$. 
The first factor  of expression (\ref{the_first_term})
(namely, $\langle\langle j_xj_y\rangle\rangle$)
is closely related to the Drude Peak
$D=\langle\langle JJ\rangle\rangle/N$ where $J=\sum_n j_n$ is the
 total current operator. 

From the discussion of the Drude Peak\cite{Kohn} we know
that under OBC's the Drude Peak is zero 
[because $D$ can be written as the response to a static twist 
which can be removed by a gauge transformation of the form
$\exp(i\sum_n nS_n^z)$
(see Ref.\ \onlinecite{Kohn})
] whereas
it is non-zero for periodic boundary conditions (PBC's).
In our case $\langle\langle j_xj_y\rangle\rangle$ is zero under OBC's
[use a gauge transform $\exp(i\sum_{n>y} S_n^z)$].
Under PBC's we find, because of translational invariance 
and because of the continuity equation, that $\langle\langle
j_xj_y\rangle\rangle$ does depend neither on $x$ nor $y$.
This implies $D=N\langle\langle j_xj_y\rangle\rangle$.
Since the Drude peak is finite (at least for the models we are
interested in) 
we conclude that expression
(\ref{the_first_term})
 vanishes even under PBC's in the Thermodynamic limit.
So we obtain
\begin{equation}g={\rm Re}[(z\hbar)^{-1}]{\rm Re}\int_0^\infty e^{izt}
\langle[j_x(t),j_y]\rangle dt.
\label{intbypart1}\end{equation}

Restarting from Eq.\ (\ref{defg}) we may---again by partial
integration---arrive at another formula.
\begin{eqnarray*}
g
&=&\!{\rm Re}\left[\frac{i}{\hbar}\Bigl\{-\langle [P_x,P_y]\rangle -(iz)
\int_0^\infty\!\! e^{izt}\langle[P_x(t),P_y]\rangle dt\Bigr\}\right]. \\
\end{eqnarray*}
The first term in the square brackets does not contribute---as the
potentials $P_x$ and $P_y$ commute---and if we 
restrict ourselves to 
${\rm Re}z=0$ we obtain:
\begin{equation}g=-{\rm Im}z{\rm Im}\left(
\frac{1}{\hbar}\int_0^\infty e^{izt}\langle[P_x(t),P_y]\rangle dt\right).
\label{intbypart2}
\end{equation}
The latter formula is especially useful for MC-simulations as it
allows the computation of the conductance in terms
of the diagonal $S^z$-$S^z$-correlators (under OBC's).

According to its  definition as it is given by Eq.\ (\ref{defg})
the conductance might  in principle depend on the actual choice of the
 positions of the voltage drop $y$ and
the current measurement $x$.
Here, we point out that  in the limit $z\to 0$
this is not the case.
In a rather general situation one can show (using the continuity equation)
that the right hand side 
of Eq.\ (\ref{defg}) gives the same result for any choice of $x$ and $y$.
(see Appendix \ref{proofxy}.)

\begin{figure}
\epsfig{file=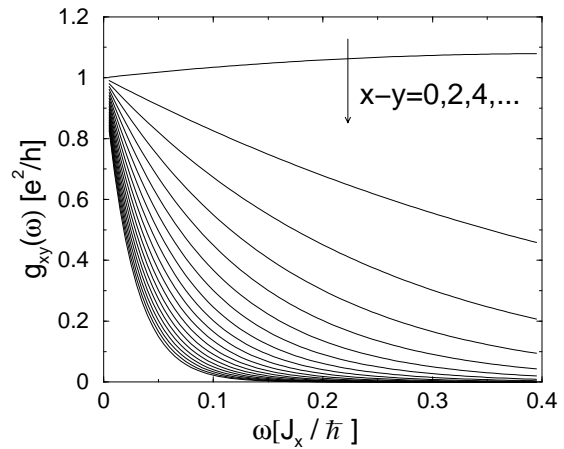,width=0.4\textwidth,angle=0}
\caption{The conductance of the $xy$ chain at $T=0.01$
versus frequency on the imaginary axis
for various distances ($x-y=2n, n\in{\mathbb N}$) between voltage drop ($y$)
and current measurement ($x$). 
}
\label{gxy}
\end{figure}

It is  instructive to consider the free fermion case {\it en d\'etail}.
We denote the---formal---dependence
 on $x$ and $y$ by corresponding subscripts. 
Of course, in a translational invariant system $g_{xy}$ depends only on
the difference $x-y$. 
The conductance $g$ as a function of $\omega={\rm Im}z$ (here and in the sequel ${\rm Re}z=0$)
is plotted for the free fermion case in Fig.\ \ref{gxy}. One sees that
the conductance in the limit $\omega\to 0$ approaches the universal
value $e^2/h$. Here we emphasize that a spatial separation of
voltage drop and current measurement leads to an exponential
decrease in $g(\omega)$ at small $\omega$.
Therefore, we will restrict our attention to  the cases $|x-y|\leq 1$ for the
rest of the paper. 

\subsection{Technical details of the MC-method}
We now turn to some technical details of our simulations.
The second formula for the conductance, Eq.\ (\ref{intbypart2}),
 lends itself to a 
study with
Monte Carlo simulations (it requires OBC's).
 At the Matsubara frequencies $\omega_M=2\pi
M(\beta\hbar)^{-1},\; M\in{\mathbb N}$ we may use the 
equivalent expression 
$$g(\omega_M)=-\omega_M/\hbar {\rm Re}\,\int_0 ^{\hbar\beta} \langle P_x
P_y(i\tau)\rangle e^{i\omega_M \tau}d\tau.$$
 We employ a standard QMC method (SSE) to compute the
conductance. Since $P_x$ is diagonal in the $S^z$-Basis, the simulation
of $\langle H^kP_x H^{L-k}P_y\rangle$ can be easily performed with the
help of the SSE.\cite{Sandvik,Dorneich,Sandvik2}
Here, $L$ is the approximation order. 
One may simply obtain $\langle P_xP_y(i\tau)\rangle$
as a linear combination of the terms  $\langle H^kP_x H^{L-k}P_y\rangle$
with binomial weight factors
$B(\tau,k)$. We found it convenient
to assume a Gaussian distribution for the $B(\tau,k)$
instead of a binomial one, 
because the former is easier to evaluate.
The error that we introduce by this replacement is smaller
than the statistical error if  $L>100$.
 (Note that in our simulations $L$ is typically of the order of
 $10^4-10^5$.)

What remains to be done in order
 to get $$g(\omega_M)=-\omega_M/\hbar\int_0^{\hbar\beta} \cos(\omega_M
\tau)\langle P_xP_y(i\tau)\rangle d\tau
$$ is an integration in the final step.
We performed it with the Simpson rule and a grid of 800 $\tau$-values.

We are now left with the standard problem of extrapolating
$g(\omega)$ from the Matsubara frequencies $\omega_M$ to $\omega=0$.
Unfortunately, the spacing of the Matsubara frequencies is linear
in $T$, so our method becomes unstable when we increase $T$.

\begin{figure}[b]
\epsfig{file=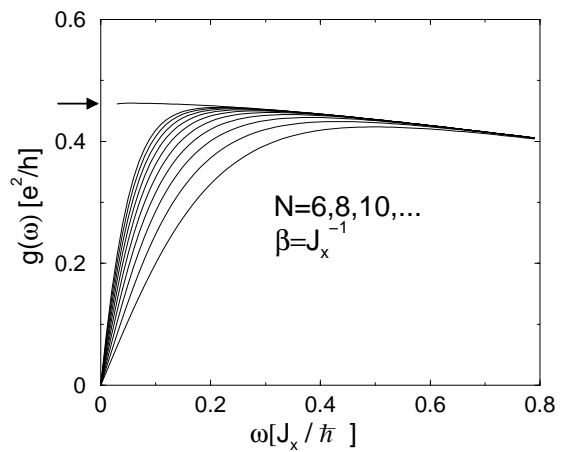,width=0.4\textwidth,angle=0}
\caption{Conductance $g$ as a function of $\omega$. The arrow indicates
 the result of the Thermodynamic limit. }
\label{cond}
\end{figure}

To see that an application of our MC-method makes only sense at low
temperatures we compare it to a  simpler method: exact diagonalization.
Fig.\ \ref{cond} shows $g(\omega)$ for various system sizes $N$ 
at $T=J_x/k_B$.
One sees that convergence with $N$ is rather fast (at high temperatures).
Hence, one can determine $g(\omega)$
for $\omega > J_x/\hbar$ with exact diagonalization.
 To compute $g(\omega)$ at some $\omega\leq J_x/\hbar$
 with MC-methods one needs
to work at a temperature $T<J_x/(2\pi k_B)$, and even then exact diagonalization
is preferable as it yields $g(\omega)$ in a continuous interval rather
than on a discrete set of points. Hence, at high temperatures
the MC-method is inferior to a simple exact diagonalization.

In our simulation we make one ``MC-sweep'' between two measurements
 which consists
of one diagonal update and
 several loop-updates\cite{Sandvik} between two measurements.
We are able to run approximately $10^5$ sweeps.

\subsection{Test with Jordan-Wigner }
If $J_z=0$ then $g(\omega)$ can be exactly evaluated---for arbitrary system
size---not only at $\omega=0$ (see below).

\begin{figure}
\epsfig{file=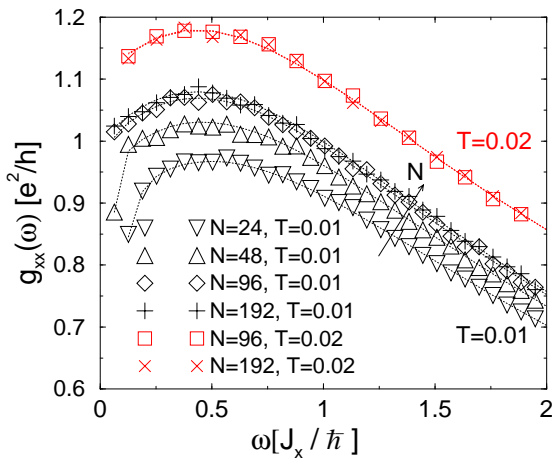,width=0.4\textwidth,angle=0}
\caption{Monte Carlo data (symbols) in comparison with the exact
 Jordan-Wigner result 
 (dotted lines). The curves for $T=0.02$ are offset by $0.1$ for clarity.
(We use OBC's.)}
\label{condmc}
\end{figure}

We can exploit this fact in two regards:
Firstly, we test  our MC-method by comparing it with the exact curve
obtained by Jordan-Wigner. The result can be seen
in Fig.\ \ref{condmc}.

Secondly, we can test which frequencies (and hence which temperatures)
we need such that a linear extrapolation can be carried out
 without introducing a larger error than the statistical one.
One sees also which system sizes are needed to determine $g(\omega)$
without detectable finite size error.
Our conclusion is that our method works for $T<0.02$.
At $T\approx 0.01$ a system size of $N\approx 200-300$ is appropriate.
\section{The conductance in various systems }

\subsection{Results for a dimerized Jordan-Wigner-chain}
One may derive a simple analytical result for the conductance in the free
fermion case. Here, we consider the slightly more difficult case---but also
more interesting as the system has a gap\cite{Orignac}---of a dimerized
chain with magnetic field $B$, i.e., $J_z=0$ and the hopping parameter 
alternates:
$(J_x)_{2n,2n+1}=J_1$ and $(J_x)_{2n+1,2n+2}=J_2$.
The energy dispersion is given by 
$$E_k^\pm=B\pm\sqrt{J_1^2+J_2^2+2J_1J_2\cos(2k)}\,/\,2
$$
and we assume a positive dispersion $E_k=E_k^+$
if  $k\in [0,\pi/2]\cup[3\pi/2,2\pi]$ and $E_k=E_k^-$ else. The
gap $E_g$ is $E_g=E_{\pi/2}^+-E_{\pi/2}^-$ at $k=\pm\pi/2$.
The evaluation of the formula for the conductance Eq.\ (\ref{defg})
 is in principle straightforward.
(see Appendix \ref{Apniceres}.)
One obtains the compact result:
\begin{eqnarray}g=&&
 \frac{e^2}{2h}\big[\tanh(E_0\beta/2)-\tanh(E_\pi\beta/2)\nonumber\\
&&-\tanh(E_{\pi/2}^+
\beta/2)+\tanh(E_{\pi/2}^-\beta/2)\big].
\label{niceresult}
\end{eqnarray}
In the case of zero magnetic field this reduces to:
$$g={ \frac{e^2}{h}\big[\tanh(E_0\beta/2)-\tanh(E_{\pi/2}^+
\beta/2)\big]}. $$
The $T=0$-value of the conductance as a function of
magnetic field is quantized:
It is zero if $|B|$ is smaller than the zero field gap $E^+_{\pi/2}$ or larger
than the zero field band width $E_0$, and
it is $1$ between these values, and  precisely
at these values it is $1/2 $ (all values in units of $e^2/h$).

\subsection{Comparison with the Apel-Rice-formula}
\begin{figure}
\epsfig{file=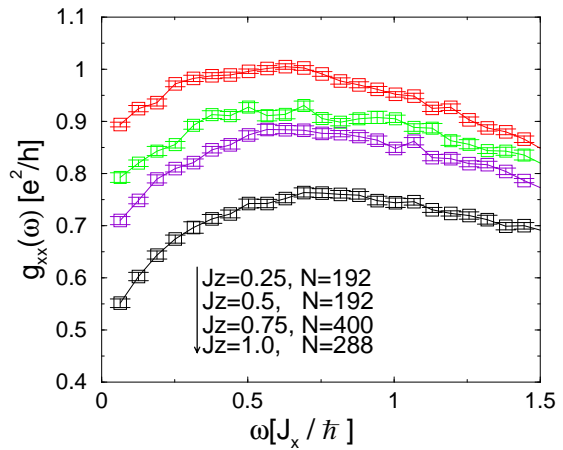,width=0.4\textwidth,angle=0}
\caption{Monte Carlo data (symbols) for various 
interaction strengths at $T=0.01J_x/k_B$ (using OBC's and
$2\cdot 10^5$ Monte-Carlo sweeps). Shown is the conductance
as a function of imaginary frequency.}
\label{QmcgJz}
\end{figure}

\begin{figure}
\epsfig{file=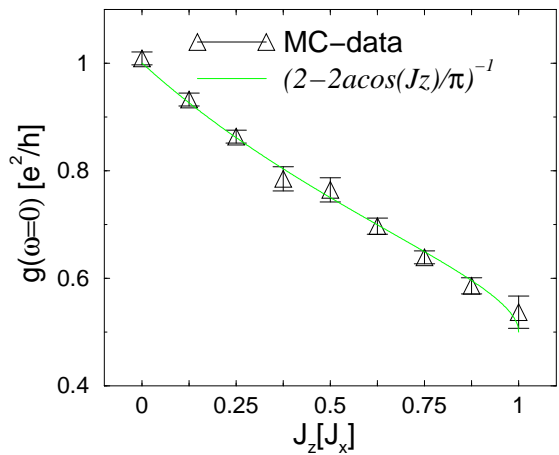,width=0.4\textwidth,angle=0}
\caption{Conductance of the $xxz$-spin chain at $T=0.01J_x/k_B$
in comparison with the exact result by Apel and Rice 
(\ref{ApelRice}) for a Luttinger Liquid. 
The Monte-Carlo data are obtained by
extrapolating the results of Fig.\ \ref{QmcgJz} to $\omega=0$.
}
\label{apelrice}
\end{figure}

At low temperatures the $xxz$-chain can be described by a Luttinger
liquid. For this model the conductance was first obtained by Apel and
Rice in the eighties:\cite{ApelRice,Giamarchi}
\begin{equation}\label{ApelRice}
g_{\rm ApelRice}=\frac{e^2}{h}\frac{\pi}{2(\pi-\theta)},
\end{equation}
where $\cos\theta=J_z/J_x$. 
This formula may be derived from Eq.\ (\ref{intbypart2}) if we use concrete
expressions for $\langle S^z_nS_m^z(i\tau)\rangle$ which are available
from conformal field theory\cite{Affleck}.
Fig.\ \ref{QmcgJz} shows our QMC-results for
$g(\omega)$ on the imaginary axis for the $xxz$-model,
in Fig.\ \ref{apelrice} we display a comparison, as a function
of $J_z$, between the
$g(\omega=0)$ extrapolated from Fig.\ \ref{QmcgJz},
and the exact Bosonization result, Eq.\ (\ref{ApelRice}).

We note, that the statistical error of the QMC-results presented in
Fig.\ \ref{apelrice} does not increase much with the parameter $J_z$.
This is due to our using the ``directed loops'' as described in Ref.\
\onlinecite{SandSyl}. The choice for transition probabilities which was
proposed there makes the SSE-algorithm more effective. Here the
improvement is remarkable.

\subsection{System with one impurity }
Now, we consider the Hamiltonian
$$H=H_{xxz}+B_{\rm Imp}S^z_{N/2},$$
i.e., we add an impurity in the middle of the system.
This model was first studied in a paper by Kane and
Fisher.\cite{KaneFish}
Later, this kind of model received considerable attention by other
authors.\cite{Leung, Weiss,Fendley,Tsvelik,Qin}
By an RG approach Kane and Fisher
found that the perturbation $B_{\rm Imp}$ is  relevant
for repulsive interactions (i.e., $J_z>0$) .
This means that at zero $T$ the chemical potential anomaly ``cuts''
the system into two halves, such that  the conductance is zero.
This result cannot be directly confirmed by Monte Carlo methods
since these are necessarily finite-temperature methods.

\begin{figure}
\epsfig{file=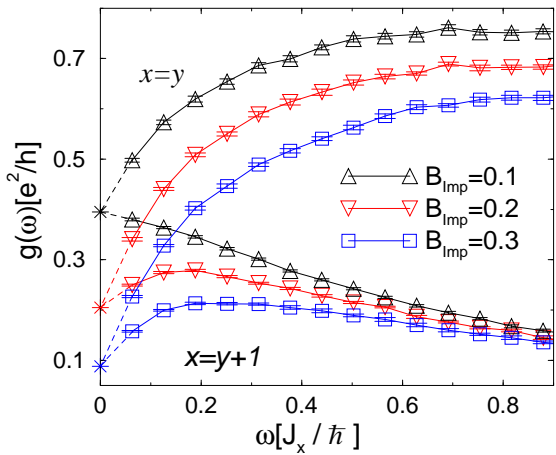,width=0.4\textwidth,angle=0}
\caption{Conductance of the Heisenberg-chain ($N=400$ sites) 
with one impurity at $T=0.01J_x/k_B$ 
for different Impurity strengths and positions of voltage drop and current 
measurement (using OBC's). A possible extrapolation to $\omega=0$ is
 proposed by the dashed lines.
}
\label{KFG2}
\end{figure}

Fortunately,  the scaling behavior (with temperature) 
of the conductance is also known. For $K=1/2$ 
one may  derive an exact formula for 
the conductance by a refermionization technique:\cite{Weiss,GogNerTsve}
\begin{equation}\label{WES}
g=e^2/h\left[1-\frac{B_{\rm Imp}^2}{2\pi^2 T}\psi^\prime
\left(1/2+\frac{B_{\rm Imp}^2}{2\pi^2T}\right)\right]\end{equation}
where $\psi$ is the Digamma function. So we can compare our
MC-data once again with an exact result.
In Fig.\ \ref{KFG2} we present two different QMC-results for the
conductance on the imaginary axis of the Heisenberg-chain with one impurity,
for different impurity strengths.

\begin{figure}
\epsfig{file=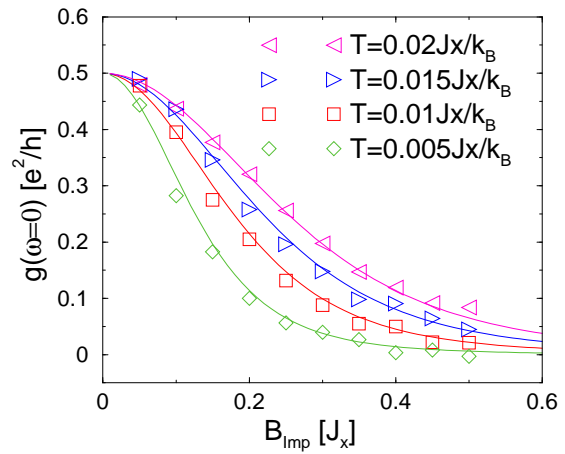,width=0.4\textwidth,angle=0}
\caption{Conductance of the Heisenberg-chain with one impurity at
 various
temperatures ($N=800$ for $T=0.005J_x/k_B$ and $N=400$).
MC-data (symbols) are drawn
in comparison with the exact formula (solid lines)
by  Weiss et al.\cite{Weiss} for a Luttinger
Liquid at the isotropic point $K=\frac{1}{2}$ (i.e., $J_z=J_x$). 
}
\label{sassetti}
\end{figure}

For the upper set of curves in Fig.\ \ref{KFG2}
the position of voltage drop and the
position of the current measurement are $x=y=N/2$; 
and for the lower set of curves they are $x=N/2,\;y=N/2-1$.
The curves are not as smooth as the ones in Fig.\ \ref{QmcgJz},  
so we use a  quadratic fit from the first 
three Matsubara frequencies instead of a linear extrapolation
to estimate $g(\omega=0)$. We also note that the 
curves with  $x-y=1$ are better suited for extrapolation than those with
$x=y$ because the slope at
$\omega=0$ is smaller.
The statistical error is less than one percent.
The result from the extrapolation is given in Fig. \ref{sassetti}
along with the exact curves from Eq.\ (\ref{WES}).
We used system sizes of $N=400$ for $T\geq 0.01J_x/k_B$ 
and $N=800$ for $T=0.005J_x/k_B$. We performed $2\cdot 10^5$ MC-sweeps.
The error bars are smaller than the symbol size, so the error that
we see in the figure is mainly due to our extrapolation method [and to
possible logarithmic finite-temperature-corrections to Eq. (\ref{WES})]. We see that the
 quadratic fit tends to underestimate the correct value.

\subsection{Inhomogeneous systems}

Several years after the publication of the Eq.\ (\ref{ApelRice}) by Apel
and Rice it was generally agreed upon\cite{Safi,MaslovStone} 
that it does not reflect the (correct) physical behavior one 
would encounter in an experimental realization.
Experiments are never performed on a closed system but on 
one coupled to reservoirs which make it possible for the particles 
to leave and enter the system.
These reservoirs can be modeled by attaching two leads
consisting of infinite non-interacting spin half-chains to our model. 

The complete Hamiltonian reads then:
\begin{equation}
H=H_{xxz}(J_z=0) +\!\!\!\! \sum_{n=(N-N_I+1)/2}^{(N+N_I-3)/2}\!\!\!\! 
S_n^zS_{n+1}^z,
\label{H_leads}
\end{equation}
i.e., the interaction is  confined to a small region in the middle
consisting of $N_I$ sites.
This approach has been followed by many authors, e.g.\ Refs.\
\onlinecite{Molina,EggerGrab,Meden,Eggeretal}.
Generally, the presence of leads yields a conductance which is
independent of $J_z$,\cite{Safi,MaslovStone} namely,
$$g=e^2/h$$ in sharp contrast to Eq.\ (\ref{ApelRice}).
The non-interacting semi-chains---which we call leads---play the r\^ole of
reservoirs. 

We note that the behavior of $g(\omega)$ depends on the parity of $N_I$
a fact 
which was already reported in Ref.\ \onlinecite{Molina}.
(A similar effect was also found in the Hubbard model.\cite{Oguri})
In the following we will only consider the case of $N_I$ odd.
A detailed discussion of $N_I$ even/odd and
a comparison with Ref.\ \onlinecite{Molina} will be presented elsewhere.

\begin{figure}
\epsfig{file=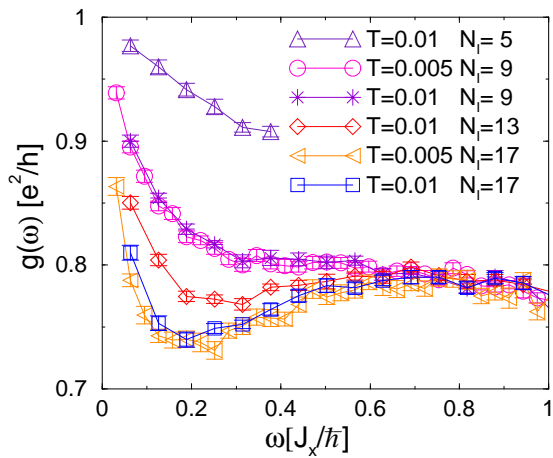,width=0.4\textwidth,angle=0}
\caption{Conductance, in units of $e^2/h$, of the Heisenberg chain  
with non-interacting leads, see Eq.\ (\ref{H_leads}). The system size 
is $N=320$. The values for temperature $T$ and size of the interacting
 region $N_I$ are given.
(We use OBC's, $10^5$ MC-sweeps, and $x=y$ for  the
voltage drop.)
}
\label{leads}
\end{figure}

The ``natural'' choice for current measurement and voltage drop would be
at the two ends of the interacting region. But here is caution advised.
From Fig.\ \ref{gxy}---which is again for the free Fermion case---one 
learns two things: Firstly, $g$ does not depend on the choice of
 $x$ and $y$. Secondly, if $x$ and $y$ are  some distance apart,
convergence with $\omega$ becomes slow, hence one needs to go to lower
$T$
and larger system sizes if one wants to extract $g(\omega=0)$ reliably.
A simple phenomenological explanation for this
 is the following:
If the place of the measurement is far from the voltage drop the
particles
have to travel a long distance and hence one has to wait a long time,
before one can determine $g$.
We can now place both the voltage drop and our current measurement at
the middle of the system, but this should not help much.
The problem is that the particles still have to travel a long distance
until they see the leads. Hence $g(\omega)$ will be unaffected by the
introduction of leads if $\omega$ is sufficiently large.
This expectation is confirmed by the QMC-data presented in
Fig.\ \ref{leads}.
In this figure the exponential decay of $g(\omega)$ at
	  small $\omega$ is apparent, and
the curves illustrate clearly  that this decay is induced by the length
	  scale $N_I$---because it becomes stronger with increasing $N_I$.
It is this decay that prevents us from discussing larger $N_I$. 
   If we increase $N_I$ the decay becomes stronger, hence we need to
	  evaluate $g(\omega)$ for more
	  (and smaller) frequencies in order to extract $g(\omega=0)$
	  reliably.
           But smaller frequencies are only available at smaller temperatures.
          As we cannot decrease $T$ much below $0.01J_x/k_B$
          we  restrict ourselves to $N_I<20$.
        If we considered a system with $N_I=200$ (at $T=0.01J_x/k_B$)
 we would not see any difference from a system
	  without leads, because the difference occurs at small $\omega$.

The data presented in Fig.\ \ref{leads} clearly indicates an
upturn of $g(\omega)$ for $\omega\to0$, indicating that
$g(\omega=0)$ is unaffected by the interaction in the low-temperature
limit. Our result may, however, not be totally convincing,
since we can only analyze relatively small
interacting regions. One might argue that the enhanced conductance is
not due to the leads but simply to the reduced ``mean'' 
interaction---which is close to zero as only few sites interact.
To invalidate this argument we considered another model. 
We have performed QMC-simulation of a system where we
attach a lead only at {\it one} side such that we obtain
a chain which is non-interacting in one half and
interacting in the other. For this system we found no deviation
at all imaginary frequencies
from the situation where the interacting region extends over the
whole chain, even though there are as many interacting as
interaction-free bonds.

\subsection{Spin-Hamiltonian with third-nearest-neighbor interaction}
Monte Carlo simulations allow the inclusion of long-range hopping, and
thus breaking the integrability of the pure $xxz$-Hamiltonian,
as long as the resulting system is not frustrated. We thus consider
a Hamiltonian with a third-nearest-neighbor interaction:
\begin{eqnarray} \nonumber
H\ =\ H_{xxz}
 &+&J_{x3}\sum_n \big[\,(S_n^+S_{n+3}^-+S_n^-S_{n+3}^+)/2\\
&&\ \ +\,J_{z}/J_xS_n^zS_{n+3}^z\,\big].
\label{Hamilhop3}
\end{eqnarray}
For simplicity we assumed that the anisotropy is independent of the
hopping range (i.e., $J_{z3}=J_{x3}J_{z}/J_{x}$). In this context
we emphasize, that the long-range hopping in the spin
system does not transform under Jordan-Wigner to a long-range hopping in
a fermionic system but to a more complicated four-sites operator.

In general, adding a new term to the Hamiltonian changes the current
operator which is defined via the continuity-equation,
$\nabla j:=(j_{n+1}-j_n)=i[S_{n+1}^z,H]/\hbar$.
In a one-dimensional system with OBC's the continuity-equation
is solved however by the relation $ j_x=i[H,P_x]/\hbar$
(see Sec.\ \ref{subSec_def_g}) such that 
Eq.\ (\ref{intbypart2}) still applies.
Nonetheless, it is useful to look at the current operator for this case.
It reads:
$$j_n=j_{n,1}+j_{n,3}+j_{n-1,3}+j_{n-2,3},$$ 
where
$j_{n,k}=iJ_xe\left(S_n^+S_{n+k}^--S_n^-S_{n+k}^+\right)/(2\hbar).$
If we compare it with the current operator of the $xxz$ chain
we see that the long range hopping $J_{x3}$ gives rise to 
three additional terms which are analogous to the first one.

\begin{figure}
\epsfig{file=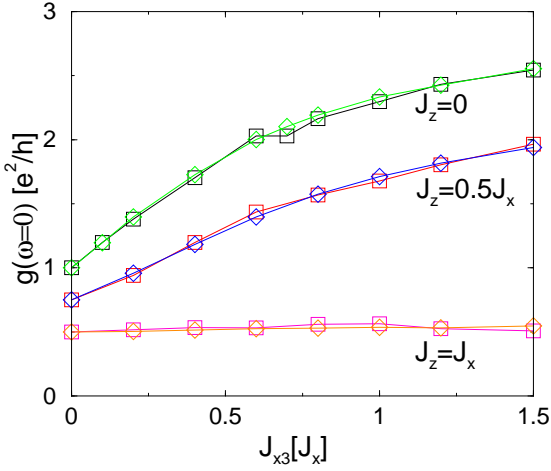,width=0.4\textwidth,angle=0}
\caption{Conductance of 
the Hamiltonian  Eq. (\ref{Hamilhop3}) (with
400 sites, $T=0.01$) as a function of $J_{x3}$ for various anisotropies.
squares: quadratic fit from the first three
 Matsubara frequencies; diamonds: linear fit from the first six
 frequencies. (We use OBC's, $10^5$ MC-sweeps, and $x=y+1$ for  the
voltage drop.) }
\label{hop3}
\end{figure}

We compute the conductance as a function of 
the hopping amplitudes $J_{x3}/J_{x}$ and present the result in Fig.\ 
\ref{hop3}. If $J_{x3}=0$ the conductance is of
course given by the Apel-Rice-result Eq.\ (\ref{apelrice}).
However, if $J_{x3}>>J_x$ we may eventually neglect the
nearest-neighbor-hopping-term such that we end up with three uncoupled
chains.
Thus, we conclude that the conductance will grow towards
 three times the Apel-Rice-result when we increase $J_{x3}$.
>From the figure we see that the crossover between these two values is
shifted to smaller values of $J_{x3}$ when the anisotropy is increased.
(In fact, at the isotropic point the increasing of $g$ is barely visible.)

In conclusion we have developed a QMC-technique which allows
the evaluation of the DC-conductance for a wide range of
non-frustrating quantum-spin chains at low but finite temperatures.
We have presented several stringent tests for this technique, 
like the Kane-Fisher scaling for the conductance 
through an impurity in a Luttinger-liquid.

\begin{appendix}
\section{Proof that the right hand side of Eq.\ (\ref{defg}) is independent of $x$ and $y$}
\label{proofxy}
Here we provide a general argument which relies on the (physical)
assumption that some linear response functions
are finite in the Thermodynamic limit.

{{\bf Theorem:}\quad{\it In a  Spin system 
the conductance
$g_{xy}\equiv g$ does not depend on $x$ and $y$  if the linear response of an operator $S_n^z$ to a
perturbation $P_m$  is bounded in the Thermodynamic limit.}}

{\bf Proof:}

First we consider
 $\lim_{z\to 0}z\int_0^\infty e^{izt}\langle
[S^z_n(t),P_m]\rangle dt.$
This expression corresponds to a plateau value of the response function
$\varphi(t):= i\langle [P_m,S_n^z(t)]\rangle$, i.e.,
$$\lim_{z\to 0}z\int_0^\infty e^{izt}\left\langle
[S^z_n(t),P_m]\right\rangle dt=\lim_{t\to \infty}\varphi(t).$$
The response function may be written in the following way
(by Kubo's identity)\cite{ZMR}
$$\varphi(t)=\beta(P_m,iLe^{iLt}S_n^z) =:\dot\Phi(t).$$
Here $(\cdot,\cdot )$ is the Mori scalar product 
[for operators $A$ and $B$: $\beta(B,A)=
\int_0^\beta{\rm Tr} B^\dag\exp(-\tau H)A\exp((\tau-\beta) H)d\tau/{\rm Tr} \exp(-\beta H)$] and 
$L$ is the  Liouville operator.
To prove that $\varphi(t)\to 0$ as $t\to\infty$,
it is sufficient to show that
 $\Phi$ as a function of $t$ is  bounded. 

But this is just the assumption that we made in the statement of the theorem
because $(S_n^z(t),P_m)$ represents the linear response of the operator $S^z_n$
to the perturbation $P_m$.

Finally, we can prove our main assertion.
We want to show $g(x,y)=g(x^\prime,y^\prime)\;\forall x,y,x^\prime,y^\prime$.
This follows easily from
$g(x,y)=g(x+1,y)\;\forall x,y$ and $g(x,y)=g(x,y+1)\;\forall x,y$.
We will only consider only the second equality [the proof of the first
equality is analogous by the symmetric structure of  Eq.\ (\ref{intbypart2})].

 Using Eq.\ (\ref{intbypart2}) and our previous result
we see:
$$g(x,y)-g(x,y+1)=
\lim_{z\to 0}{\rm Im }z\int_0^\infty\!\! e^{izt}\langle
 [S_{y+1}^z(t), P_x]\rangle dt=0. $$
\begin{flushright}Q.E.D.\end{flushright}

\section{Derivation of Eq.\ (\ref{niceresult})}
\label{Apniceres}
 For eigenvalues $E_n$ and $E_k$ the respective one-particle-eigenstates
will be denoted by $|n\rangle$ and $|k\rangle$, the annihilation
 operators by $c_n$ and $c_k$, and the occupation numbers
by  $n_n$ and $n_k$.

For the current operator we find:
 $N\langle k,j_xk\rangle =ev_k:=e/\hbar\frac{dE}{dk}$
and $\langle -k,j_xk\rangle=0$.

In a free Fermion system one can derive a simple expression for
 $\langle A(i\omega_M)B\rangle=\int_0^\beta d\tau e^{i\omega_M\tau}\langle
AB(i\tau)\rangle$ [with $i\omega_M=\omega+i\delta=2\pi iM/(\hbar\beta)$]
when $A$ and $B$ are one-particle operators (i.e., $A=\sum_{n,k}
 A_{nk}c_n^\dag c_k$):

$$\langle AB\rangle(i\omega_M)= \sum_{n\neq k} 
\frac{A_{nk}B_{kn}}{i\hbar\omega_M+E_n-E_k}f_{kn}$$
where $f_{kn}=
\frac
{\sinh(\beta(E_n-E_k)/2)}{2\cosh(\beta E_n/2)\cosh(\beta E_k/2)}=
\left(
\langle n_k\rangle -\langle n_n\rangle\right).$
$\langle A(z=\omega+i\delta)B\rangle$ may then be obtained by 
analytic continuation. In our case $A$ and $B$ are local current operators;
for the conductance we use Eq.\ (\ref{intbypart1}):

\begin{eqnarray*}
g&=& \frac{1}{\omega}{\rm Im}\sum_{n\neq k} 
\frac{\langle n,j_x k\rangle\langle k,j_y n\rangle}
{i\hbar\omega_M+E_n-E_k}f_{kn} \\
&=& -\frac{1}{\omega}\sum_{n\neq k}\langle n,j_x k\rangle\langle k,j_y n\rangle 
\frac{\delta}{(\hbar\omega+E_n-E_k)^2+\delta^2}f_{kn}. 
\end{eqnarray*}
For  $\delta\to 0$ $$\frac{\delta}{(\hbar\omega+E_n-E_k)^2+\delta^2}\to
\pi\delta(\hbar\omega+E_n-E_k).$$ In the continuum limit we replace
$\sum_kN^{-1}\to\int \frac{dk}{2\pi}=\int \frac{dE_k}{2\pi|\hbar v_k|}  $

Performing the integration over the variable $n$ 
and then taking $\omega\to 0$  yields 
\begin{eqnarray*}g
&=& \frac{e^2}{2\hbar}\int \frac{dk}{2\pi}
|\hbar v(k)|\frac{\beta/2}{2\cosh^2(\beta
E_k/2)}\\
&=&\frac{e^2}{\hbar}\int_{-\pi}^\pi \frac{dk}{2\pi}
\frac{\hbar v_k{\rm sgn}(k)\beta/2}{4\cosh^2(\beta
E_k/2)} =\frac{e^2}{h}\int_{k=0}^{k=\pi}\frac{dE\beta/2}{2\cosh^2(\beta
E_k/2)} \\
 &=&\frac{e^2}{2h}\int_{E_{\pi/2}^+\beta/2}^{E_0\beta/2}
 dx\frac{1}{\cosh^2(x)}+\frac{e^2}{2h}\int_{E_{\pi}\beta/2}^{E_{\pi/2}^-\beta/2}
 dx\frac{1}{\cosh^2(x)} \\
&=& \frac{e^2}{2h}[\tanh(E_0\beta/2)-\tanh(E_\pi\beta/2) \\
&&-\tanh(E_{\pi/2}^+
\beta/2)+\tanh(E_{\pi/2}^-\beta/2)].
\end{eqnarray*}

Note that the conductance does not depend on the energy dispersion but
only on  the band width $E_0$ and the gap $E_{\pi/2}^+$.

\end{appendix}


\begin{thebibliography}{aa}

\bibitem{Cas95} H.~Castella, X.~Zotos, and P.~Prelov{\v{s}}ek, 
                Phys. Rev. Lett. {\bf 74},  972 (1995). 

\bibitem{Zot96} X.~Zotos and P.~Prelov{\v{s}}ek, 
                Phys. Rev. B {\bf 53},  983  (1996).

\bibitem{Zot97} X.~Zotos, F.~Naef, and P.~Prelov{\v{s}}ek, 
                Phys. Rev. B {\bf 55}, 11029 (1997).

\bibitem{Alv02} J.V. Alvarez, C. Gros,  
                Phys. Rev. Lett. {\bf  88}, 077203 (2002);
                Phys. Rev. B {\bf 66}, 094403 (2002).

\bibitem{Imry} Y.\ Imry, {\it Introduction to Mesoscopic Physics,}
Oxford University Press (New York 1997).

\bibitem{GogNerTsve} A.\ Gogolin, A.\ Nersesyan, and A.\ Tsvelik, {\it
	Bosonization and Strongly Correlated Systems}, Camebridge
	University Press, Camebridge 1998.

\bibitem{ApelRice} W.\ Apel and T.M.\ Rice, Phys. Rev. B {\bf 26}, 7063
	(1982).

\bibitem{KaneFish} C.L.\ Kane and M.P.A.\ Fisher, Phys. Rev. Lett. {\bf 68},
	1220 (1992); Phys. Rev. B {\bf 46}, 15233 (1992). 

\bibitem{Marston} M.A.\ Cazalilla and J.B.\ Marston, ``Time-dependent
	density-matrix renormalization group'' Phys. Rev. Lett. {\bf 88},
	256403 (2002); {\it ibid.} {\bf 91}, 049702, 2003 reply to comment
         H. G. Luo, T. Xiang, and X.Q. Wang {\it ibid.},
          {\bf 91} 049701, (2003).


\bibitem{Molina} R.A.\ Molina, D. Weinmann, R. A. Jalabert,  G.-L. Ingold, and  J.-L. Pichard Phys. Rev. B {\bf 67}, 235306, (2003).


\bibitem{Moon} K.\ Moon, H.\ Yi, C.L.\ Kane, S.M.\ Girvin, and M.P.A.\ Fisher,
	Phys. Rev. Lett. {\bf 71}, 4381 (1993).

\bibitem{Leung} K.\ Leung, R.\ Egger, and C.H.\ Mak, Phys.  Rev. Lett. {\bf 75},
	3344 (1995).

\bibitem{Sushkov} O.P.\ Sushkov, Phys. Rev. B {\bf
	64}, 155319 (2001).

\bibitem{Mescholl} V.\ Meden and U.\  Schollwoeck, Phys. Rev. B {\bf 67}, 193303 (2003).


\bibitem{Sandvik} A.W.~Sandvik,
                {\it ``Stochastic series expansion method with
                operator-loop update''},
                Phys. Rev. B {\bf 59}, R14157 (1999).


\bibitem{louis}  K.\ Louis and C.\ Gros, Phys. Rev. B {\bf 67}, 224410 (2003).

\bibitem{Kohn} W.\ Kohn, Phys. Rev. {\bf 133}, A171 (1964).

\bibitem{Dorneich} A.\ Dorneich and M.\ Troyer, Phys. Rev. E {\bf 64},
	066701 (2001).

\bibitem{Sandvik2} A.W.~Sandvik and J.~Kurkij\"arvi, Phys. Rev. B {\bf
	43}, 5950 (1991); A.W.\ Sandvik, J. Phys. A {\bf 25}, 3667
	(1992).

\bibitem{Orignac} E.~Orignac, R.~Chitra, and R.~Citro, Phys. Rev. B {\bf
	67}, 134426 (2003). condmat/0211633

\bibitem{Giamarchi}
T.~Giamarchi and H.J.~Schulz, Phys. Rev. B {\bf 37}, 325 (1988).

\bibitem{SandSyl} O.F.~Sylju{\aa}sen and A.W.~Sandvik,  Phys. Rev. E {\bf
	66}, 046701 (2002).



\bibitem{Affleck} S.~Eggert, I.~Affleck, and M.~Takahashi,
	Phys. Rev. Lett. {\bf 73}, 332 (1994).

\bibitem{Weiss} U.~Weiss, R.~Egger, and M.~Sassetti, Phys. Rev. B {\bf 52}, 16707 (1995). 

\bibitem{Fendley} P.\ Fendley, A.W.W.\ Ludwig, and H.\ Saleur,  Rev. Lett. {\bf 74}, 3005 (1995).


\bibitem{Tsvelik} A.\ Tsvelik, J. Phys. A {\bf 28}, 625-L (1995).

 
\bibitem{Qin} S.~Qin, M.~Fabrizio, and L.~Yu, Phys. Rev. B {\bf 54},
	R9643 (1996); see also a discussion in 
V.\ Meden, P.\ Schmitteckert, and N. Shannon, Phys. Rev. B {\bf 57}, 8878 (1998).


\bibitem{Safi} I.\ Safi and H.J.\ Schulz,  Phys. Rev. B {\bf 52},
	R17040 (1995).

\bibitem{MaslovStone} D.L.~Maslov and M.~Stone, Phys. Rev. B {\bf 52},
	R5539 (1995).


\bibitem{EggerGrab} R.~Egger and H.~Grabert, Phys. Rev. Lett. {\bf 77},
	538 (1996); ibidem {\bf 80}, 2255(E) (1998).


\bibitem{Meden} V.~Meden, S.~Andergassen, W.~Metzner, U.~Schollwoeck,
	and K.~Schoenhammer,  condmat/0303460 (unpublished).

\bibitem{Eggeretal} R.~Egger, H. Grabert, A. Koutouza, H. Saleur, and F. Siano, Phys.\ Rev.\ Lett.\ {\bf 84}, 3682
	(2000).

\bibitem{Oguri} A.~Oguri, Phys.\ Rev.\ B {\bf 59}, 12240 (1999).


\bibitem{ZMR} D.~Zubarev, V.~Morozov, and G.~Roepke,  
              {\it ``Statistical mechanics of nonequilibrium processes''},
              Akademie Verlag, Berlin 1997.

\end{thebibliography}
\end{document}